\documentclass{aastex62}
\usepackage{rotating}
\usepackage[flushleft]{threeparttable}

\received{May 10, 2018}
\accepted{July 3, 2018}
\submitjournal{ApJ}

\shorttitle{Chemical inhomogeneities in the Pleiades}
\shortauthors{Spina et al.}

\begin{document}

\title{Chemical inhomogeneities in the Pleiades: signatures of rocky-forming material in stellar atmospheres}

\correspondingauthor{Lorenzo Spina}
\email{lorenzo.spina@monash.edu}

\author[0000-0002-9760-6249]{Lorenzo Spina}
\affil{Monash Centre for Astrophysics, School of Physics and Astronomy, Monash University, VIC 3800, Australia}
\affiliation{Universidade de S\~ao Paulo, IAG, Departamento de Astronomia, Rua do Mat\~ao 1226, S\~ao Paulo, 05509-900 SP, Brasil}

\author{Jorge Mel\'{e}ndez}
\affiliation{Universidade de S\~ao Paulo, IAG, Departamento de Astronomia, Rua do Mat\~ao 1226, S\~ao Paulo, 05509-900 SP, Brasil}

\author{Andrew R. Casey}
\affil{Monash Centre for Astrophysics, School of Physics and Astronomy, Monash University, VIC 3800, Australia}

\author{Amanda I. Karakas}
\affil{Monash Centre for Astrophysics, School of Physics and Astronomy, Monash University, VIC 3800, Australia}

\author{Marcelo Tucci-Maia}
\affiliation{N\'{u}cleo de Astronom\'{i}a, Universidad Diego Portales, Av. Ej\'{e}rcito 441, Santiago, Chile}



\begin{abstract}

The aim of Galactic archaeology is to recover the history of our Galaxy through the information encoded in stars. An unprobed assumption of this field is that the chemical composition of a star is an immutable marker of the gas from which it formed. It is vital to test this assumption on open clusters, group of stars formed from the same gas. Previous investigations have shown that unevolved stars in clusters are chemically homogeneous within the typical uncertainties of these analysis, i.e. 15$\%$ of the elemental abundances. Our strictly differential analysis on five members of the Pleiades allows us to reach precisions of 5$\%$ for most elements and to unveil chemical anomalies within the cluster that could be explained by planet engulfment events. These results reveal that the evolution of planetary systems may alter the chemical composition of stars, challenging our capability of tagging them to their native environments, and also paving the way for the study of planetary architectures and their evolution, through the chemical pattern of their host stars.

\end{abstract}

\keywords{stars: abundances ---  stars: chemically peculiar --- stars: solar-type --- open clusters and associations: individual (Pleiades) }


\section{Introduction} \label{sec:intro}

Open clusters are groups of stars formed within the same nebula and by the same material. For this reason, unevolved stars in an open cluster are expected to have an identical chemical content, which reflects the initial composition of the cloud. Like a DNA profile, one could use the individual chemical patterns of stars that are not in clusters today to trace back their common origin to their native formation sites \citep{Freeman02}. This approach -- commonly called ``chemical tagging'' -- is a powerful tool for galactic archeology, which aims at recovering the remnants of the ancient building blocks of the Milky Way (e.g., clusters, super-clusters, moving groups) that are now dispersed, and reconstructing their star formation history. In fact, the main impulse behind the large-scale spectroscopic surveys of the current decade (e.g., APOGEE, Gaia-ESO, GALAH; \citealt{Holtzman15,Gilmore12,DeSilva15}) has been the acquisition of large and homogeneous sets of spectroscopic data from different environments within the Galaxy to trace its history in space and time. 

A vital condition for chemical tagging is that stellar clusters are chemically homogeneous, as they constitute the left-overs of the star-forming clouds and the starting point of stellar populations. At what level can this assumption be supported by the observations? To date, several important tests have confirmed that unevolved stars in open clusters are chemically homogeneous at the level of the measurement uncertainties, typically 0.05 dex or larger \citep{DeSilva06,Mitschang13,BlancoCuaresma15}. Nevertheless, underlying chemical inhomogeneities may limit our capability of chemical tagging at higher precision. 

There are different astrophysical sources that may be able to imprint chemical inhomogeneities among the unevolved cluster members, violating the founding assumption of chemical tagging. For example, a nearby supernova that have polluted an inefficiently mixed star-forming cloud would have produced scatter in newly synthesised elements \citep{Reeves72}. Nevertheless, \citet{Bovy16} has recently put tight constraints on the $initial$ abundance scatter in three open clusters (i.e. $\lesssim$0.2 dex), indicating that the gas and the dust in protostellar clouds is very well mixed before most of the stars formed. On the other hand, it also has been proposed that the presence of planetary systems and their evolution can alter the pristine atmospheric compositions shortly after the stellar formation \citep{Melendez09,Spina15}. The great diversity among the exo-planetary systems discovered so far \citep{Winn15} suggests that dynamical processes can lead planets to migrate inwards or outwards and could allow them to acquire large eccentricities \citep{Baruteau16}. Various processes acting within gaseous disks, or dynamical planetary systems, can drive the orbital decay of planets ending in their accretion onto the host star \citep{Jackson17}. For instance, it is likely that a massive exoplanet migrating inward would induce other innermost planets to move into unstable orbits \citep{Mustill15}. When this material enters the star, it is rapidly dissolved in the stellar envelope, altering the star's chemical pattern in a way that mirrors the composition observed in rocky objects: with refractory elements (i.e., those with condensation temperature T$_{\rm cond}$$>$1000~K) being more abundant than volatiles \citep{Chambers10}. Studies on binary systems (e.g., \citealt{TucciMaia14,Biazzo15,Ramirez15,Teske15,Teske16,Saffe16,Saffe17,Oh18}) and open clusters \citep{Spina15} have demonstrated that differences can exist in the composition of stars belonging to the same association and that these chemical anomalies are typically larger for the most refractory elements.

In order to further test the chemical homogeneity of unevolved stars formed by the same cloud, we observed five members of the Pleiades, a 120 Myr old open cluster \citep{Kharchenko05}. These are ``adolescent'' stars: they are old enough not to limit a spectroscopic analysis (e.g., high extinction, strong accretion, veiling, fast rotation), but they are still young so that their planetary systems have formed only $\sim$100 Myr ago. The newly born planets are still cleaning their orbits from gas and dust left by the circumstellar disk and this viscous process makes planetary migration extremely likely. In addition, these stars are not too old so mixing mechanisms between the stellar layers had not quenched the consequences of pollution by refractory elements from the planetary system \citep{Theado12}.

\section{Observations and analysis} \label{sec:analysis}
The targets have been selected from the catalog of Pleiades candidate members compiled by \citet{Stauffer07}. The stars have B-V colours within 0.62-0.73, similar to the Solar values (i.e., [B-V]$_{\odot}$=0.65; \citealt{Ramirez12}) and with projected rotational velocities v$sin$i$\lesssim$10 km~s$^{-1}$ (\citealt{Glebocki05}; see Table~\ref{parameters}). 
We observed these stars with the Ultraviolet and Visual Echelle Spectrograph (UVES; \citealt{Dekker00}) on the Very Large Telescope of the European Southern Observatory. The observations have been performed with a resolving power R$\sim$75,000 and a wavelength coverage between 330 - 680 nm. Thanks to the brightness of our targets (V$<$11.5), we achieved signal-to-noise ratios (S/N) within 350-500 pixel$^{-1}$ at 600 nm, with a median of 400 pix$^{-1}$. Further spectra including the oxygen triplet region at 777~nm have been acquired through the High Resolution Echelle Spectrometer (HIRES; \citealt{Vogt94}) on the Keck-I Telescope. These additional spectra have a R$\sim$50,000 and S/N within 200-300 pixel$^{-1}$ at 777 nm, with a median S/N of 250 pixel$^{-1}$. Solar spectra have been acquired both through UVES (S/N$\sim$400 pixel$^{-1}$) and HIRES (S/N$\sim$250 pixel$^{-1}$) adopting the same spectrograph's setups used for the cluster members. 

Equivalent widths (EWs) of 85 Fe I and 15 Fe II lines were measured in the UVES spectra using IRAF's $\tt{splot}$ task, adopting the same approach detailed in our previous works \citep{Bedell14,Spina16}. The EW measurements are processed by the qoyllur-quipu (q2) code \citep{Ramirez14} that automatically estimates the stellar parameters (effective temperature T$_{\rm eff}$, surface gravity log g, metallicity [Fe/H], and microturbulence $\xi$) by iteratively searching for the three equilibria: excitation, ionisation, and the trend between the iron abundances and the reduced equivalent width log$_{\rm 10}$$\big[$$\frac{EW}{\lambda}$$\big]$. 

The observed objects are twin stars, meaning that they have similar atmospheric parameters. This allowed us to perform the spectral analysis through a strictly line-by-line differential approach relative to the Solar spectrum, in order to minimise the effects of the systematic uncertainties in the stellar models and in the atomic transition parameters (e.g., \citealt{Bedell17,Nissen15,Nissen16,Liu14,Liu16a,Liu16b,Spina18}) For this procedure we assumed the nominal solar parameters, T$_{\rm eff}$=5777 K, log g=4.44 dex, [Fe/H]=0.00 dex and $\xi$ =1.00 km s$^{-1}$ \citep{Cox00}. The iterations are executed with a series of steps starting from a set of initial parameters (i.e., the solar parameters) and employing the Kurucz (ATLAS9) grid of model atmospheres \citep{Castelli04}. In each step the abundances are estimated using MOOG (version 2014, \citealt{Sneden73}). The errors associated with the stellar parameters are evaluated by the code taking into account the dependence between the parameters in the fulfilment of the three equilibrium conditions \citep{Epstein10,Bensby14}. The atmospheric parameters resulting from our analysis, listed in Table~\ref{parameters}, confirm the twin nature of our targets, as they are within a restricted range of values:  $\pm$231~K, $\pm$0.05~dex, $\pm$0.08~dex, for T$_{\rm eff}$, log~g, and [Fe/H], respectively. The typical uncertainties are $\sigma$$_{\rm Teff}$$=$23 K, $\sigma$$_{\rm logg}$$=$0.04, dex$\sigma$$_{\rm [Fe/H]}$$=$0.016 dex, $\sigma$$_{\rm \xi}$$=$ 0.04 km~s$^{-1}$.

According to the Gaia DR2 \citep{GaiaDR2}, the stars have parallaxes comprised within 7.16 and 7.68 mas (listed in Table~\ref{parameters}), which are values expected for members of the Pleiades \citep{Kharchenko05}.


In the stellar spectra, we measured 126 absorption lines of atomic transitions for 18 elements in addition to iron: C, O, Na, Mg, Al, Si, S, Ca, Sc, Ti, V, Cr, Mn, Co, Ni, Cu, Zn and Y. Titanium and chromium were detected in both the neutral and first ionised states. The oxygen lines were measured in HIRES spectra, while UVES spectra were used to measure all the other EWs. Using MOOG through the q2 code and adopting the stellar parameters determined as described above, we calculated the abundance corresponding to each line detected in the spectra of the five stars and the Sun. Through the $\tt{blends}$ driver in MOOG and adopting the line list from the Kurucz database, the q2 code corrected the abundances of V, Mn, Co, Cu, and Y for hyperfine splitting effects. It also calculated the non-LTE corrections \citep{Ramirez07} for the oxygen abundances as a function of the stellar parameters. Then, the final abundances have been determined through a strict line-by-line differential approach using the Sun as the reference star, taking the averages over the measured abundances for all lines of each element. The error budget associated with each elemental abundance has been obtained summing in quadrature the standard error on the mean and the propagated effects of the uncertainties on the stellar parameters. When only one line of an individual element is detected, the standard error was estimated by repeating the EW measurement five times with different assumptions on the continuum setting. The resulting elemental abundances and their uncertainties are listed in Table \ref{abu_Sun}, where the values reported in brackets are the number of lines used for the abundance determination of a specific element.

\begin{table}
\centering
\begin{threeparttable}
\caption{Stellar rotational velocities, parallaxes and atmospheric parameters.}
\label{parameters}
\medskip
\begin{tabular}{c|cc|cccc}
\hline
Star & v$sin$i & $\pi$ & T$_{\rm eff}$ & $log$~g & [Fe/H] & $\xi$ \\
 & [km s$^{-1}$] & [mas] & [K] & [dex] & [dex] & [km s$^{-1}$] \\
\hline
HIP17020 & 9.6$\pm$1.1 & 7.16$\pm$0.07 & 6049$\pm$19 & 4.58$\pm$0.03 & 0.054$\pm$0.015 & 1.43$\pm$0.03 \\
HII~250 & 5.9$\pm$0.8 & 7.68$\pm$0.07 &  5818$\pm$21 & 4.56$\pm$0.04 & 0.022$\pm$0.018 & 1.31$\pm$0.04 \\
HII~514 & 10.5$\pm$1.0 & 7.36$\pm$0.05 &  5854$\pm$32 & 4.57$\pm$0.05 & 0.039$\pm$0.023 & 1.42$\pm$0.05 \\
HD282962 & 5.1$\pm$1.3 & 7.35$\pm$0.04 &  5922$\pm$22 & 4.53$\pm$0.04 & $-$0.023$\pm$0.017 & 1.38$\pm$0.03 \\
HD282965 & 10.3$\pm$0.4 & 7.16$\pm$0.05 &  6014$\pm$21 & 4.58$\pm$0.04 & 0.022$\pm$0.015 & 1.39$\pm$0.03 \\
\hline
\end{tabular}
 \begin{tablenotes}
  \small
  \item Note. Rotational velocities v$sin$i are from \citet{Glebocki05}. Parallaxes $\pi$ are from \citet{GaiaDR2}.
    \end{tablenotes}
  \end{threeparttable}
\end{table}

\begin{sidewaystable}
\centering
\caption{Differential abundances relatively to the Sun.}
\label{abu_Sun}
\begin{tabular}{cccccc|c}
\hline
Element & HIP17020 & HII 250 & HII 514 & HD282962 & HD282965 & c$_{j}$ \\
 & [dex] & [dex] & [dex] & [dex] & [dex] & [dex]\\
\hline
C I & $-$0.096$\pm$0.022 (2) & $-$0.083$\pm$0.021 (2) & $-$0.106$\pm$0.033 (1) & $-$0.084$\pm$0.026 (3) & $-$0.110$\pm$0.023 (2) &  $-$0.095$\pm$0.011\\
O I & $-$0.006$\pm$0.031 (3) & 0.021$\pm$0.028 (3) & 0.045$\pm$0.046 (3) & 0.033$\pm$0.033 (3) & 0.029$\pm$0.037 (3) &  0.021$\pm$0.015\\
Na I & $-$0.040$\pm$0.039 (2) & $-$0.077$\pm$0.046 (2) & $-$0.034$\pm$0.061 (2) & $-$0.102$\pm$0.022 (2) & $-$0.082$\pm$0.042 (2) &  $-$0.070$\pm$0.016 \\
Mg I & 0.005$\pm$0.017 (5) & $-$0.039$\pm$0.017 (5) & 0.006$\pm$0.024 (5) & $-$0.077$\pm$0.016 (5) & $-$0.017$\pm$0.016 (5) & $-$0.025$\pm$0.010 \\
Al I & $-$0.036$\pm$0.012 (1) & $-$0.047$\pm$0.011 (1) & $-$0.057$\pm$0.019 (1) & $-$0.091$\pm$0.020 (2) & $-$0.081$\pm$0.011 (1) &  $-$0.063$\pm$0.009 \\
Si I & $-$0.002$\pm$0.013 (7) & $-$0.021$\pm$0.006 (12) & 0.026$\pm$0.008 (7) & $-$0.068$\pm$0.006 (13) & $-$0.015$\pm$0.013 (7) & $-$0.017$\pm$0.007 \\
S I & $-$0.022$\pm$0.036 (2) & $-$0.020$\pm$0.018 (2) & $-$0.033$\pm$0.030 (2) & $-$0.018$\pm$0.021 (2) & 0.000$\pm$0.018 (2) &  $-$0.012$\pm$0.011\\
Ca I & 0.105$\pm$0.018 (9) & 0.045$\pm$0.019 (10) & 0.081$\pm$0.027 (8) & 0.013$\pm$0.019 (10) & 0.063$\pm$0.016 (9) & 0.063$\pm$0.011 \\
Sc II & $-$0.008$\pm$0.021 (4) & $-$0.055$\pm$0.019 (5) & $-$0.002$\pm$0.026 (4) & $-$0.118$\pm$0.021 (5) & $-$0.018$\pm$0.017 (4) &  $-$0.039$\pm$0.012\\
Ti I & 0.057$\pm$0.018 (10) & 0.014$\pm$0.023 (14) & 0.031$\pm$0.032 (8) & $-$0.029$\pm$0.022 (14) & $-$0.005$\pm$0.019 (8) &  0.015$\pm$0.012\\
Ti II & 0.049$\pm$0.018 (7) & $-$0.008$\pm$0.019 (7) & 0.012$\pm$0.033 (4) & $-$0.070$\pm$0.020 (8) & 0.014$\pm$0.022 (6) &  0.001$\pm$0.012\\
V I & 0.083$\pm$0.030 (5) & 0.038$\pm$0.025 (7) & 0.043$\pm$0.033 (5) & 0.020$\pm$0.029 (7) & 0.017$\pm$0.022 (6) & 0.039$\pm$0.014 \\
Cr I & 0.053$\pm$0.021 (8) & 0.022$\pm$0.022 (11) & 0.008$\pm$0.034 (6) & $-$0.002$\pm$0.020 (12) & 0.008$\pm$0.020 (8) & 0.022$\pm$0.012 \\
Cr II & 0.066$\pm$0.022 (3) & 0.021$\pm$0.019 (3) & --- & 0.001$\pm$0.022 (4) & 0.054$\pm$0.029 (3) & 0.038$\pm$0.013 \\
Mn I & $-$0.094$\pm$0.023 (4) & $-$0.076$\pm$0.022 (5) & $-$0.087$\pm$0.033 (4) & $-$0.133$\pm$0.034 (5) & $-$0.122$\pm$0.028 (4) & $-$0.101$\pm$0.013 \\
Fe I & 0.054$\pm$0.015 (61) & 0.022$\pm$0.018 (84) & 0.039$\pm$0.023 (48) & $-$0.023$\pm$0.017 (82) & 0.022$\pm$0.015 (57) &  0.024$\pm$0.010\\
Fe II & 0.057$\pm$0.018 (9) & 0.017$\pm$0.023 (14) & 0.044$\pm$0.032 (9) & $-$0.021$\pm$0.022 (14) & 0.019$\pm$0.025 (7) &  0.024$\pm$0.012\\
Co I & $-$0.075$\pm$0.020 (5) & $-$0.078$\pm$0.017 (6) & $-$0.053$\pm$0.025 (3) & $-$0.120$\pm$0.024 (8) & $-$0.079$\pm$0.021 (2) &  $-$0.081$\pm$0.011\\
Ni I & $-$0.021$\pm$0.014 (15) & $-$0.062$\pm$0.015 (17) & $-$0.032$\pm$0.021 (15) & $-$0.118$\pm$0.014 (17) & $-$0.060$\pm$0.018 (15) &  $-$0.060$\pm$0.009\\
Cu I & $-$0.073$\pm$0.011 (1) & $-$0.124$\pm$0.013 (1) & --- & $-$0.113$\pm$0.048 (2) & $-$0.135$\pm$0.014 (1) & $-$0.112$\pm$0.008\\
Zn I & $-$0.124$\pm$0.009 (1) & $-$0.126$\pm$0.018 (2) & $-$0.151$\pm$0.018 (1) & $-$0.143$\pm$0.012 (2) & $-$0.163$\pm$0.010 (1) &  $-$0.142$\pm$0.006\\
Y II & 0.137$\pm$0.020 (5) & 0.093$\pm$0.020 (2) & 0.108$\pm$0.030 (5) & 0.060$\pm$0.022 (4) & 0.092$\pm$0.023 (5) & 0.100$\pm$0.013 \\
\hline
\end{tabular}
\end{sidewaystable}

\section{The chemical inhomogeneities in the Pleiades} \label{sec:results}
The elemental abundances obtained through our analysis indicate that chemical inhomogeneities exist within the five Pleiades members. In Table~\ref{inhomogeneities} we list, for each star, the observed scatter $\sigma$$_{obs}$ and the expected scatter $\sigma$$_{exp}$ in the chemical patterns relatively to the median composition of the cluster. The first is defined as the quadratic sum of differential abundances between the star and the median cluster, while the second is the quadratic sum of the abundance uncertainties. We also report in Table~\ref{inhomogeneities} the corresponding $\chi$$^{2}$=$\sigma$$^{2}_{obs}$/$\sigma$$^{2}_{exp}$. Typically, any chemically anomalous star is identifiable by a $\chi$$^{2}$$>$1. The values listed in Table~\ref{inhomogeneities} indicate that the Pleiades is chemically inhomogeneous up to the level of $\sigma$$_{int}$/$\sqrt{N-1}$=0.04 dex, where $\sigma$$^{2}_{int}$=$\sigma$$^{2}_{obs}$-$\sigma$$^{2}_{exp}$ and N is the number of detected species calculated for the most chemically diverse star, HD282962 ($\chi$$^{2}$=2.9). Another chemically anomalous star is HIP17020 ($\chi$$^{2}$=1.6), which has a mildly discrepant chemical pattern at the level of $\sigma$$_{int}$/$\sqrt{N-1}$=0.02 dex. The star HII~514 has a chemical composition that is marginally consistent with the cluster median (i.e., $\chi$$^{2}$=0.95). Finally, the other two members have a composition in agreement with the cluster median (i.e., $\chi$$^{2}$$\sim$0.3). 

\begin{figure*}
\centering
\includegraphics[width=10cm]{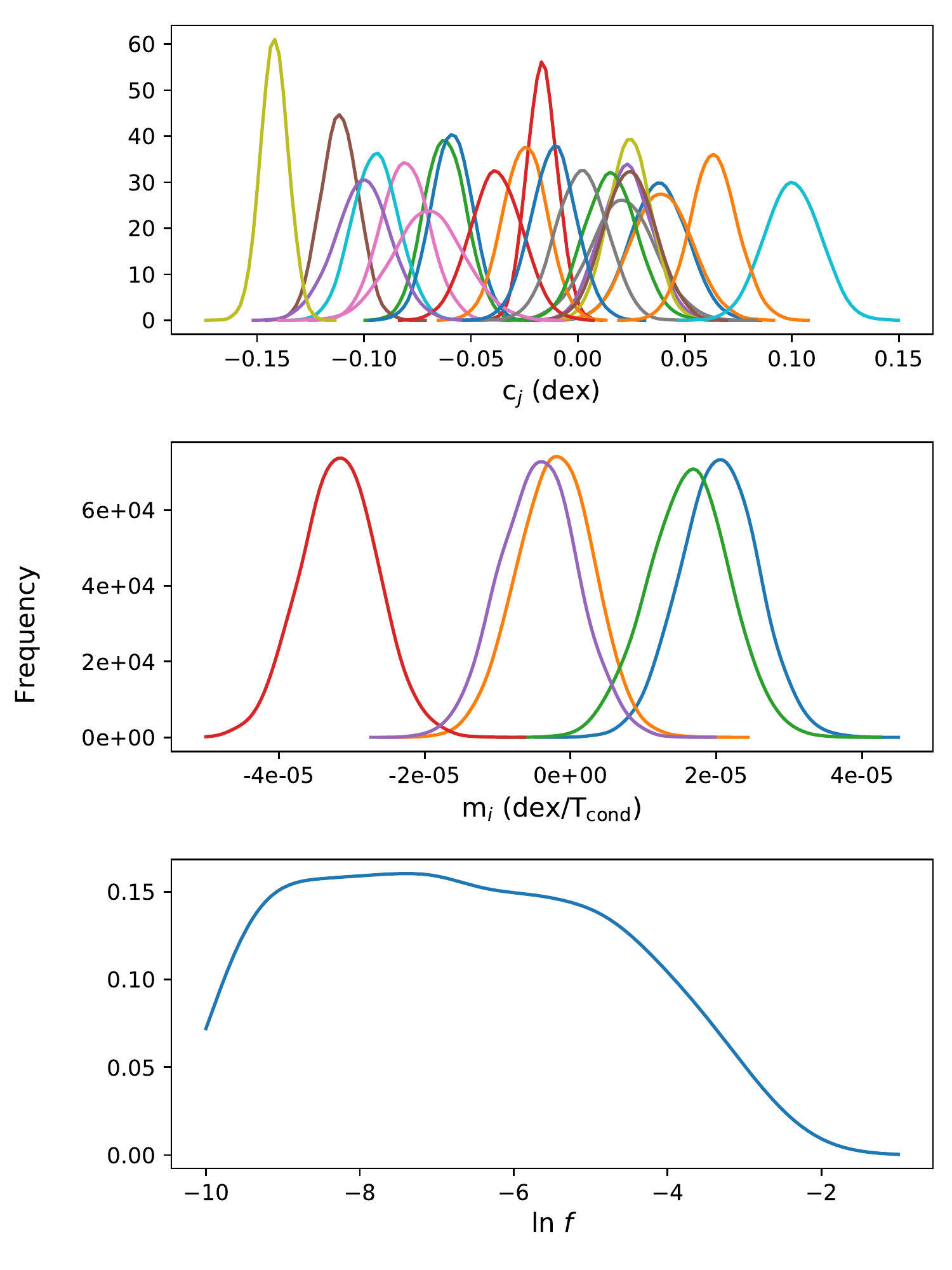}
\caption{The kernel distributions calculated for the parameters c$_{j}$, m$_{i}$ and $f$ employed in the fitting procedure. \label{fig_trace}}
\end{figure*}

A detailed inspection of the stellar compositions can provide insights on the possible origin of the chemical inhomogeneities in the Pleiades. With this aim, we simultaneously modelled the cluster abundances and allowed for star-to-star variations as a function of the dust condensation temperature T$_{\rm cond}$ for a Solar-system composition gas. Namely, we fitted the abundance values in Table \ref{abu_Sun} using the No-U-Turn Sampler \citep{Hoffman11}, which is a variant of Hamiltonian Monte Carlo used by Stan \citep{Stan18}. For the procedure we adopted the following model:
\begin{equation}
Y_{ij}=c_{j}+m_{i}\times X_{j}+f\times \sigma_{ij},  \\
\end{equation}
where Y$_{ij}$ is the logarithmical abundance of the $i$-star and the $j$-element, X$_{j}$ is the condensation temperature of the $j$-element from \citet{Lodders03} and $f$ is a parameter that accounts for the possibility that the abundance uncertainties $\sigma_{ij}$ were systematically underestimated. 
We adopted a gaussian prior distribution on m$_{i}$ centred on 0 and with a sigma of 10$^{-5}$, while $\log{f}$ was allowed to uniformly range between $-$10 and 1. The marginalised kernel distributions of the c$_{j}$, m$_{i}$ and $f$ parameters are shown in the left panels of Fig.~\ref{fig_trace}. The median and the standard deviation values of the marginalised probability distributions obtained for the c$_{j}$ parameters are listed in the last column of Table~\ref{abu_Sun} and are excellent proxies of the cluster composition. The peaks (i.e., $m$$_{\rm peak}$) and the 95$\%$ (2-$\sigma$-like) credible intervals of these distributions (i.e., $m$$_{\rm low}$ and $m$$_{\rm up}$) are listed in Table~\ref{inhomogeneities}. As it is shown in the bottom panel of Fig.~\ref{fig_trace}, during the 95$\%$ of the iterations $\log{f} \leq -3$, which is negligible compared to the typical abundance uncertainties.

The left panels of Fig.~\ref{fig_trends} show the stellar chemical patterns scaled for the cluster composition (i.e., c$_{j}$) as a function of the condensation temperatures T$_{\rm cond}$. The error bars are the quadratic sum of the uncertainties in c$_{j}$ and [X/H]. The most probable m$_{i}$ values and the 95$\%$ credible intervals are over-plotted as red solid lines and shaded areas, respectively. In the right panels are shown the marginalised posterior probability distributions of the slopes m$_{i}$ resulting from our analysis.

Through these panels in Fig.~\ref{fig_trends}, we observe that the chemical patterns of HII~250 and HD282965 are in good agreement with the cluster and their slopes are consistent with zero. However, there are clear trends between the differential abundances and condensation temperature for HIP17020, HII~514, and HD282962. Notably, the chemical patterns of the most chemically anomalous stars (i.e., HD282962) is characterised by a significant negative correlation between abundances and T$_{\rm cond}$ and its range of possible slope values is inconsistent with that of any other cluster member. As it has been shown by \citet{Melendez12}, pollution by intermediate or high mass stars is not a viable explanation of such tight trends between the differential abundances and condensation temperature (see their Fig. 7). Instead these peculiar chemical patterns can be produced by a selective accretion of material from the circumstellar disk onto the star, after dust condensed in larger bodies \citep{Chambers10}. Namely, when the refractory-rich dust in protoplanetary disks condense into planetesimals that can later aggregate in planets, the circumstellar material decouples in volatile and refractory compounds. Then, a selective pollution of the stellar atmosphere from either volatiles or refractories would result in the trends discussed above. 


\begin{table}
\centering
\caption{Parameters for chemical homogeneity test.}
\label{inhomogeneities}
\medskip
\begin{tabular}{c|ccc|ccc}
\hline
Star &$\sigma$$_{\rm obs}$ & $\sigma$$_{\rm exp}$ & $\chi$$^{2}$ & m$_{\rm peak}$ & m$_{\rm low}$ & m$_{\rm up}$ \\
 & [dex] & [dex] & & [K$^{-1}$] & [K$^{-1}$] & [K$^{-1}$] \\
\hline
HIP17020 & 0.14 & 0.11 & 1.6 & 0.025 & 0.014 & 0.035 \\
HII~250 & 0.061 & 0.11 & 0.31 & $-$0.002 & $-$0.012 & 0.008 \\
HII~514 & 0.084 & 0.087 & 0.95 & 0.020 & 0.009 & 0.031 \\
HD282962 & 0.21 & 0.12 & 2.9 & $-$0.038 & $-$0.048 & $-$0.027 \\
HD282965 & 0.060 & 0.11 & 0.28 & $-$0.005 & $-$0.015 & 0.006 \\
\hline
\end{tabular}
\end{table}

\begin{figure*}
\centering
\includegraphics[width=12cm]{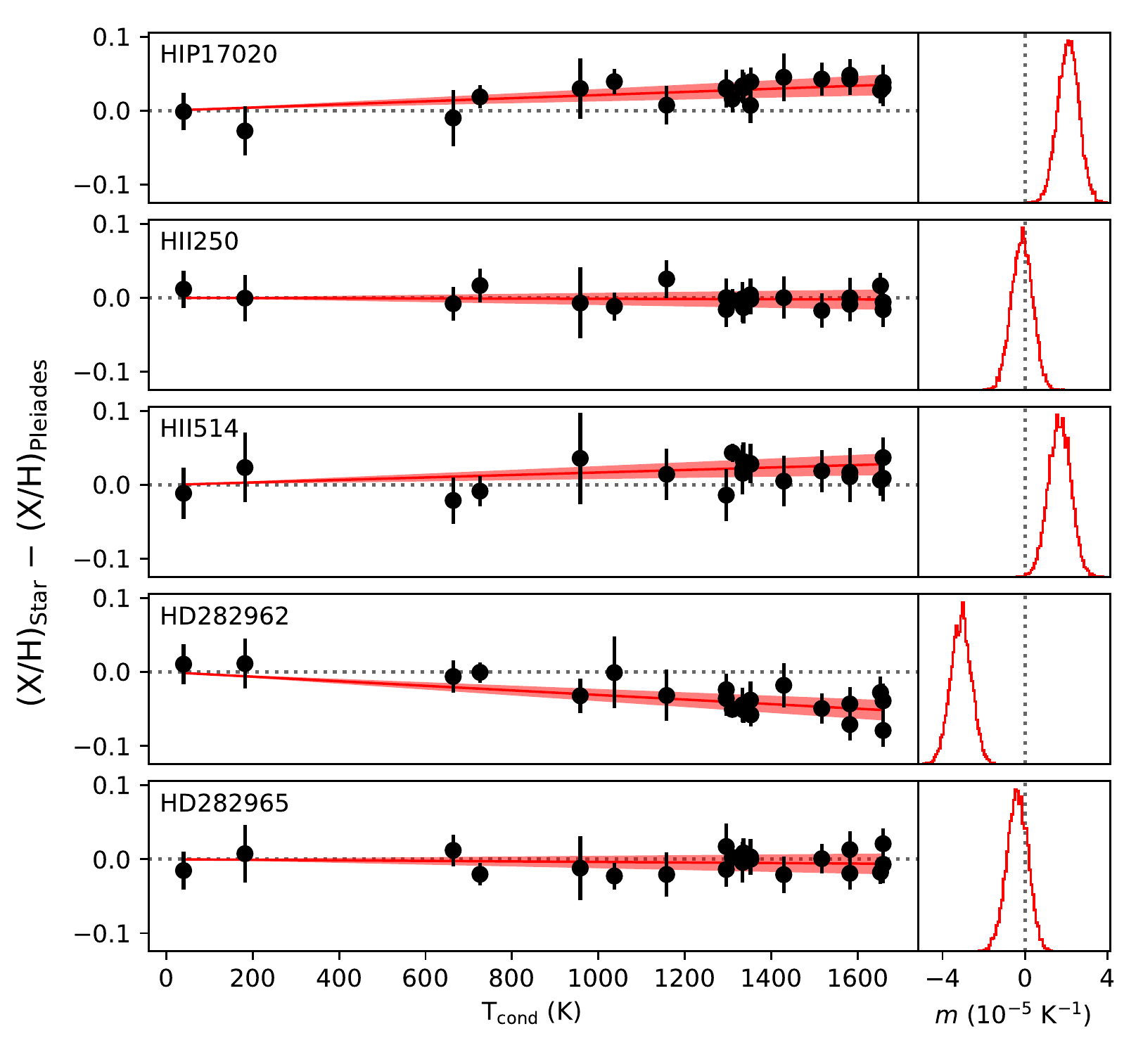}
\label{fig_trends}
\caption{The left panels show the stellar chemical patterns scaled for the cluster composition (c$_{j}$) as a function of the condensation temperatures T$_{\rm cond}$ of the elements from \citet{Lodders03}. A fitted linear regression model has been used to identify any relationship between the differential abundances and the condensation temperature. The solid red lines are the most probable $m$ slopes resulting from the fitting, while the shaded area represent the 95$\%$ credible intervals of the marginalised posterior probability distribution calculated for the slopes, which are shown in the right panels. The most probable slope and the slopes represented by the shaded areas are anchored to the point corresponding to the T$_{\rm cond}$ of the most volatile element (carbon) and the null differential abundance. \label{fig_trends}
}
\end{figure*}

\section{Fingerprints of dust condensation in stellar atmospheres}
Different scenarios can imprint chemical signatures of dust condensation similar to these observed in the stellar atmospheres of the Pleiades. 
For example, a rocky planet accreting onto a star and polluting its atmosphere would result in a selective enhancement of refractory elements \citep{Chambers10}, but it has also been proposed that an anomalous volatile-to-refractory ratio could be due to a selective accretion of volatiles after the formation of rocky planets around the star \citep{Melendez09}.

In line with a planet engulfment scenario, Fig.~\ref{fig_trends} would indicate that the five cluster members have been enriched by different amounts of rock-forming material, quantified by different $m$ slope values. This range of slopes observed in the Pleiades may reflect the heterogeneity of possible planetary systems that stars can host: from those with a very dynamical and chaotic past -- e.g., with gaseous giants on very elliptic orbits (e.g., \citealt{OToole09}) -- to the most stable systems, similar to our Solar System. Namely, during the chaotic evolution of the most dynamical systems, part of the rocky bodies may be swallowed by the star, enhancing its atmosphere in refractory elements. On the other hand, stable planetary systems can survive without any dramatic modification of their structure, being able to maintain the stellar composition unaltered. Accordingly to this view, HIP17020 would be the star that has engulfed the most rocky material from its surroundings, while HD282962 is the least enriched in refractory elements, thus it may host a more stable planetary system. This scenario opens up the intriguing possibility that the chemical composition of stars could carry information relevant to the architecture and evolution of planetary systems. 

Evolutionary models of pre-main-sequence stars allow us to infer whether the epoch of the pollution is consistent with a planet engulfment event or if the chemical atmospheres have been adulterated during the main accretion phase from the circumstellar disk. 
The stellar structure of a newly born Sun-like star comprises a convective external layer that contains more than 10$\%$ of the total stellar mass \citep{Siess00}: any amount of gas or dust falling onto the young star from its circumstellar disk would be completely diluted in such a thick convective zone, leaving the stellar composition unaltered. However, as solar-type stars evolve during their pre-main-sequence phase, they undergo a rearrangement of their internal structure, with the external convective layer retreating towards the surface. 
Therefore, it is likely that the external pollution occurred when the star was older than $\sim$20 Myr, as it is the time that the convective zone would take to became thin enough to be modified by the accretion of rocky bodies \citep{Spina15}. 
For this reason, it is plausible that the material swallowed by the stars was composed by rocky bodies (such as planets or planetesimals), rather than gas or dust that are expected to be dispersed from the disk within 10 Myr from the stellar birth \citep{Mamajek09}.

However, it is challenging to accurately predict the evolution of the internal stellar structure during the pre-main-sequence phase, as it could be affected by numerous unknown variables \citep{Kunitomo17}. For instance, recursive and vigorous accretion bursts (i.e., $\dot{\rm M}$ $\geq$ 5$\times$10$^{-4}$  M$_{\odot}$yr$^{-1}$) onto the protostar could influence the stellar evolution producing a stable and thin convective zone (2$\%$ of the total mass) earlier than 10 Myr \citep{Baraffe10}. This would allow the accretion of gas from the circumstellar disk to alter the stellar composition before and independently from planet formation. Under this assumption, the deficit of refractory elements in the atmosphere of HD282962 could be the signature of rocky planet formation \citep{Melendez09}. Namely, the refractory elements in the protostellar nebula of HD282962 would have been sequestered by the rocky bodies, while the volatiles are accreted onto the star. Conversely, the other cluster members that did not form planets were subject of an indistinctive accretion of material (volatiles plus refractories) from the disk, resulting in an enhancement in refractory elements relative to HD282962.

Analogously to the composition of HD282962, which resulted anomalously poor in refractory elements relative to other members of the Pleiades, \citet{Melendez09} have shown that the Sun has an unusual abundance pattern when compared to 11 solar twins, stars with atmospheric parameters very similar to those of the Sun. They observed that the Solar composition is characterised by a refractory-to-volatile deficiency relative to 85$\%$ of the sample. This result has been confirmed by other studies based on larger samples of stars \citep{Ramirez09,Bedell18}. A possible explanation of this puzzle is that, unlike the Sun, most of the solar twins have engulfed planets, enhancing their atmospheres of refractory elements. This hypothesis is well matched with the picture outlined by surveys for exoplanet detection. In fact, the architectures of most of the extra-solar systems discovered so far suggest that they have undergone severe processes of orbital reconfigurations in their past, while the regularity of the motion of the planets in our Solar System indicates that they formed on orbits similar to their current ones and that nothing dramatic happened to them during their lifetime \citep{Morbidelli11}. However, it is still not clear to date whether our Solar System is an extremely rare environment in the Galaxy or if the huge diversity between the Solar System and exo-planetary systems is exclusively a selection effect. As an alternative explanation, it is also possible that the chemical peculiarity observed in the Sun is a consequence of the Galactic chemical evolution \citep{Adibekyan14,Nissen15}. On the other hand, this latter hypothesis cannot explain the deficit of refractory elements found in the atmosphere of HD282962 relatively to the other members of the Pleiades, as all these stars are coeval and have formed from the same gas. Therefore, our result gives support to the possibility that signatures of planet engulfment events can truly be imprinted in the chemical patterns of stars, as also supported by planet-host binary systems composed of stellar twins, where distinct chemical compositions could be related to planets (e.g., \citealt{Oh18,Teske15,TucciMaia14})

\section{Conclusions} \label{sec:conclusions}
Regardless the precision achievable in elemental abundances, the success of chemical tagging relies on the significance of other two critical factors: the chemical diversity of open clusters and their level of chemical inhomogeneities. Studies on Galactic stellar associations have shown that the chemical scatter among stars formed in the same group may be smaller than $\sim$0.05~dex, while the groups typically differ by $\sim$0.12~dex on average for a single elements \citep{Mitschang13}. However, a higher degree of chemical overlap is expected for stellar groups that are coeval and share a similar evolutionary stage \citep{BlancoCuaresma15}. The level of chemical diversity among coeval stars has been illustrated by highly precise abundance determinations of solar twin stars: field stars within age bins of 2 Gyr have abundance variations similar to the abundance scatter observed within the members of the Pleiades \citep{Nissen15,Spina16b,Bedell18}. Therefore, the chemical inhomogeneities unveiled through this work may represent a challenge to our capability of tagging stars of similar age to different native sites.  A larger sample of stars is necessary to reliably determine the level of chemical inhomogeneities and the frequency of chemically anomalous stars within members of the same cluster. Efforts will have to be dedicated to this (e.g., by observing more stellar twins in the Pleiades or in other associations), in order to determine if these chemical anomalies constitute a significant conflict to the assumption that open clusters are chemically homogeneous.

It will also be necessary to thoroughly investigate the cause of these trends between abundances and condensation temperature. The information provided by transiting exoplanet space surveys (e.g., \citealt{Borucki10,Ricker14}) or other ground-based surveys will allow us to put on a solid ground (or rule out) any dependence between the architecture of planetary systems and the chemical composition of the host star. This is particularly necessary to understand the origin of the chemical peculiarity of the Sun \citep{Melendez09} and explore the possibility to identify stars with the best chances of hosting $analogs$ of our Solar System purely from a very careful analysis of the stellar chemical composition. This is especially urgent in an epoch that is witnessing a huge increase in the exoplanet science, as it could lay the foundations for future studies focussed on the search for Earth 2.0 and the quest for extraterrestrial life.

\acknowledgments

L.~S. and J.~M. acknowledge the support from FAPESP (2014/15706-9 and 2012/24392-2). L.~S. and A.~I.~K.  acknowledge financial support from the Australian Research Council (Discovery Project 170100521). A.~R.~C. is supported from the Australian Research Council through Discovery Grant DP160100637.
 
%

\vspace{5mm}
\facilities{VLT:Kueyen (UVES), Keck:I (HIRES)}


\software{MOOG \citep{Sneden73},  
          Qoyllur-quipu \citep{Ramirez14}, 
          Stan \citep{Stan18}
          }

\bibliography{/Users/lspina/Dropbox/papers/bibliography} 



\end{document}